\documentclass[aps,pre,twocolumn,galleyproof,superscriptaddress,groupedaddress,showpacs]{revtex4}
\usepackage{graphicx,epsf,dsfont,amssymb,verbatim}

\RequirePackage{color}
\definecolor{MyDarkGreen}{rgb}{0.02,0.60,0.06}

\begin{document}
\title{Critical phenomena on scale-free networks: logarithmic corrections and scaling functions}
\author{V. Palchykov}\email[]{palchykov@icmp.lviv.ua}
\affiliation{Institute for Condensed Matter Physics, National
Academy of Sciences of Ukraine, UA--79011 Lviv, Ukraine}
\author{C.\
von Ferber} \email[]{C.vonFerber@coventry.ac.uk}
 \affiliation{Applied Mathematics Research Centre, Coventry University,
Coventry CV1 5FB, UK}
 \affiliation{Physikalisches Institut
Universit\"at Freiburg, D-79104 Freiburg, Germany}
\author{R. Folk}\email[]{reinhard.folk@jku.at}
 \affiliation{Institut f\"ur Theoretische Physik, Johannes Kepler
Universit\"at Linz, A-4040, Linz, Austria}
\author{Yu. Holovatch}\email[]{hol@icmp.lviv.ua}
\affiliation{Institute for Condensed Matter Physics, National
Academy of Sciences of Ukraine, UA--79011 Lviv, Ukraine}
\affiliation{Institut f\"ur Theoretische Physik, Johannes Kepler
Universit\"at Linz, A-4040, Linz, Austria}
\author{R. Kenna}\email[]{R.Kenna@coventry.ac.uk}
\affiliation{Applied Mathematics Research Centre, Coventry
University, Coventry CV1 5FB, UK}
\date{Lviv, April 01, 2010}

\begin{abstract}
In this paper, we address the logarithmic corrections to the leading
power laws that  govern  thermodynamic quantities as a second-order
phase transition point is approached. For phase transitions of spin
systems on $d$-dimensional lattices, such corrections appear at some
marginal values of the order parameter or space dimension. We
present new scaling relations for these exponents. We also consider
a spin system on a scale-free network which exhibits logarithmic
corrections due to the specific network properties. To this end, we
analyze the phase behavior of a model with coupled order parameters
on a scale-free network and extract leading and logarithmic
correction-to-scaling exponents that determine its field- and
temperature behavior. Although both non-trivial sets of exponents
emerge from the correlations in the network structure rather than
from the spin fluctuations they fulfil the respective thermodynamic
scaling relations. For the scale-free networks the logarithmic
corrections appear at marginal values of the node degree
distribution exponent. In addition we calculate scaling functions,
which also exhibit nontrivial dependence on intrinsic network
properties.
\end{abstract}

\maketitle

\section{Introduction}

Scaling laws are an intrinsic feature of second-order phase
transitions. In their leading asymptotics they are power laws that
govern the behavior of the (singular part) of the free energy and of
its derivatives in the vicinity of the phase-transition point
\cite{scaling}. For a magnetic phase transition, the free energy
exhibits universal scaling in terms of its inherent variables, the
reduced temperature $\tau=|T-T_c|/T_c$ and the magnetic field $h$.
Beside the critical exponents universality manifests itself in
universal amplitude ratios and scaling functions. Moreover, a system
defined on a $d$-dimensional Euclidean space (which we will call a
{\em lattice} hereafter) becomes scale-invariant at the critical
point. Its correlation length diverges at the transition point
$\tau=0, h=0$ and the pair correlation function changes from an
exponential to a power-law decay. The leading exponents that govern
these scaling laws are related by scaling relations. These form a
cornerstone of the modern theory of critical phenomena
\cite{scaling}.

Of special interest within this theory of critical phenomena are
those situations in which the aforementioned power-laws require
logarithmic corrections \cite{logarithmic,Kenna06}. For
$d$-dimensional systems, the most prominent examples are numerous
spin models at their upper critical dimension $d_c$
\cite{ON_log,ONlr_log,phi3_log} and the $q$-state Potts model in
$d=2$ dimensions and $q = q_c=4$ \cite{Potts_log}. For spin models
the logarithmic corrections appear when the mean-field power-laws
observed for $d>d_c$  turn to non-trivial power-law dependencies at
$d<d_c$. For the Potts model, the marginal value $q_c$ separates two
different phase transition scenarios: for $q>q_c$ the transition is
of  first order, whereas for $q<q_c$ it is of second order. Another,
more subtle example is the $d=2$ Ising model with non-magnetic
impurities (see e.g. \cite{Shalaev94} and references therein).
Similar to the leading critical exponents, their logarithmic
correction counterparts have been shown to obey also a set of
scaling relations, as detailed in Ref. \cite{Kenna06}.

The situations discussed above concern systems with well-defined
Euclidean metrics and, as is clearly seen from these examples, the
notion of space dimensionality is crucial in defining the situation,
where the logarithmic corrections to scaling appear. In this paper
we want to attract attention to a different circumstance where
critical behavior requires logarithmic corrections to scaling,
namely spin models on {\em networks} or {\em random graphs}
\cite{networks_rev}. For Euclidean lattices the space dimension
implies a given coordination number ($2d$ for the $d$-dimensional
hypercube). For the networks we will consider here, these
coordination numbers (or degrees) are distributed according to a
given degree distribution. This amounts to a difference of principal
between the origin of logarithmic corrections on regular lattices
and on such networks.

The interest to study critical phenomena on complex networks is
motivated by a number of reasons \cite{Dorogovtsev08} both of
academic and practical nature. Some models on complex networks may
describe exotic phenomena (such as opinion formation in a social
network \cite{sociophysics}) as well as traditional physical objects
(e.g., integrated nanoparticle systems with complex geometry
\cite{Tadic05}). Real-life complex networks are often characterized
by a scale-free behavior: a power law decay of the node degree
distribution
\begin{equation}\label{rrr001}
P(k)\sim k^{-\lambda}.
\end{equation}
Here, $P(k)$ is the probability that an arbitrary chosen node of a
network has a degree (the number of links attached to this node)
$k$. The exponent $\lambda$ is crucial in determining the critical
behavior of different models on complex networks (see
Refs.\cite{networks_rev,Dorogovtsev08} and references therein). The
general situation is as follows: for small $\lambda < \lambda_s$ the
system is always ordered, only an infinite temperature field is able
to destroy the order. For large $\lambda>\lambda_c$ the phase
transition is described by the usual mean-field critical exponents,
whereas systems with intermediate values $\lambda_s < \lambda <
\lambda_c$ are generally described by $\lambda$-dependent critical
exponents. It is the marginal value of $\lambda=\lambda_c$ at which
the logarithmic corrections to scaling appear as has been
established for a number of classical spin models on scale-free
networks \cite{spin_on_networks}. The emergence of these corrections
signals the relevance of {\em correlations} between node degrees due
to the presence of high-degree nodes (hubs). Here, one observes a
certain similarity with the critical behavior on lattices, where the
logarithmic corrections appear at the upper critical dimension $d_c$
at which the trivial mean-field exponents turn to the non-trivial
ones due to the {\em correlations} in thermal fluctuations.

In the present work we pay special attention to the analogy between
the role of the upper critical dimension $d_c$ on a regular lattice
and the exponent $\lambda_c$ on a complex network. To this end, we
consider the field and temperature dependencies of thermodynamic
quantities that characterize the system in the vicinity of the phase
transition. The specific example we consider is a system with two
coupled order parameters on a scale-free network. This model is
widely used to describe ordering phenomena in systems with two
possible types of ordering. Physical examples are given by
ferromagnetic and antiferromagnetic, ferroelectric and
ferromagnetic, structural and magnetic ordering \cite{coupled}. In
sociophysics applications \cite{sociophysics}, one may think about
opinion formation where a coupling exists between the preferences
for a candidate and a party in an election. Recently, we have used a
Landau-like approach and a mean-field analysis to obtain the phase
diagram of this model on a complex scale-free network
\cite{Palchykov09}. In the present paper we extend this analysis to
derive the full set of critical exponents that govern the scaling
laws for the thermodynamic quantities in terms of functions of $h$
at $\tau=0$ and of functions of $\tau$ at $h=0$. A special focus of
our paper is the logarithmic-correction-to-scaling behavior. We
check the validity of existing relations for the
logarithmic-correction-to-scaling exponents and further derive new
scaling relations for exponents of logarithmic corrections, for
which these relations were so far unknown.

\section{Critical exponents and logarithmic corrections to scaling}

The behavior of a system near a second-order phase transition is
described by a number of critical exponents. The magnetization $M$,
susceptibility $\chi$ and heat capacity $C_h$ at zero external
field, $h=0$, respectively follow the power laws \cite{Privman91}
\begin{equation}\label{rrr002}
M\sim \tau^{\beta},
\end{equation}
\begin{equation}\label{rrr003}
\chi\sim \tau^{-\gamma},
\end{equation}
\begin{equation}\label{rrr004}
C_h\sim \tau^{-\alpha}.
\end{equation}
Spatial characteristics of the system, namely, the correlation
length and the correlation function, which are connected with the
linear size and the spatial dimension $d$, scale with their critical
exponents $\nu$ and $\eta$ correspondingly. The exponents connected
to the spatial structure of the lattice are not well defined for the
network. At the phase transition temperature $\tau=0$ the
dependencies of the thermodynamic characteristics on the external
field are also described by a number of critical exponents
\cite{Privman91}
\begin{equation}\label{rrr005}
M\sim h^{1/\delta},
\end{equation}
\begin{equation}\label{rrr006}
\chi\sim h^{-\gamma_c},
\end{equation}
\begin{equation}\label{rrr007}
C_h\sim h^{-\alpha_c}.
\end{equation}
The eight critical exponents listed above depend just on a few
parameters -- spatial dimension, spin dimension and symmetries of
the model. Therefore, from a knowledge of just two of the exponents
as well as the dimension, any other may be determined. Indeed, the
remaining six exponents are related via the following four scaling relations:
\begin{eqnarray}
\alpha + 2\beta + \gamma & = & 2,\label{rrr008}\\
\beta(\delta-1) & =& \gamma ,
\label{rrr009}\\
\gamma_c & = & 1 - \frac{1}{\delta}
\label{rrr010}\\
\alpha_c & = & \frac{2 + \gamma}{\beta + \gamma} - 2 .
\label{rrr011}
\end{eqnarray}
For the $d$-dimensional lattices, the behavior
(\ref{rrr002})--(\ref{rrr007}) is valid from the lower to the upper
critical dimension. Beyond the upper critical dimension, the
exponents become those predicted by the mean field approximation.
Just at the upper critical dimension one may see modifications to
the dependencies described above: there appear logarithmic
corrections \cite{Kenna06}. In the absence of an external field
($h=0$) the scaling behavior at the upper critical dimension is
\begin{equation}\label{rrr012}
M\sim \tau^{\beta}|\ln \tau|^{\hat{\beta}},
\end{equation}
\begin{equation}\label{rrr013}
\chi\sim \tau^{-\gamma}|\ln \tau|^{\hat{\gamma}},
\end{equation}
\begin{equation}\label{rrr014}
C_h\sim \tau^{-\alpha}|\ln \tau|^{\hat{\alpha}},
\end{equation}
while at the critical temperature ($\tau=0$) one finds
\begin{equation}\label{rrr015}
M\sim h^{1/\delta}|\ln h|^{\hat{\delta}},
\end{equation}
\begin{equation}\label{rrr016}
\chi\sim h^{-\gamma_c}|\ln h|^{\hat{\gamma}_c},
\end{equation}
\begin{equation}\label{rrr017}
C_h\sim h^{-\alpha_c}|\ln h|^{\hat{\alpha}_c},
\end{equation}
These hatted exponents for the logarithmic corrections are also
connected via scaling relations, and in Ref.~\cite{Kenna06} the
following formulae, which are analogous to (\ref{rrr008}) and
(\ref{rrr009}), were derived:
\begin{equation}\label{rrr018}
\hat{\beta}(\delta-1) = \delta\hat{\delta} - \hat{\gamma},
\end{equation}
\begin{equation}\label{rrr019}
\hat{\alpha} = 2\hat{\beta} - \hat{\gamma}.
\end{equation}

As it was outlined in the Introduction, we are interested in scaling
laws for the magnetic phase transition on networks with, generally
speaking, undefined Euclidean metrics. Therefore, the exponents we
will be interested in are those given by Eqs.
(\ref{rrr002})--(\ref{rrr007}) that do not involve the space
dimension $d$. The scaling relations for them are given by
(\ref{rrr008})--(\ref{rrr011}). However, only two corresponding
relations for the hatted exponents,  (\ref{rrr018}), (\ref{rrr019}),
are available in the literature so far \cite{Kenna06}. Therefore,
before we proceed further, we derive in the next section the scaling
relations for the exponents $\hat{\gamma}_c$ (\ref{rrr016}) and
$\hat{\alpha}_c$ (\ref{rrr017}) that characterize logarithmic
corrections to the field-strength dependency.

\section{New scaling relations for logarithmic corrections}\label{III}
In \cite{Kenna06} a Lee-Yang analysis was used to derive  relations
between the logarithmic-correction exponents, which are analogous to
the conventional scaling relations between the leading exponents.
Here, these considerations are extended to deal with logarithmic
corrections to the field-dependency of the susceptibility. The
Lee-Yang analysis concerns the zeros of the partition function in
the plane of  complex magnetic field. The locus of such zeros
terminates at the so-called Yang-Lee edge $r_{\rm{YL}}$, which is
temperature dependent. Following \cite{Kenna06}, we account for the
possible existence of logarithmic corrections to the scaling of the
edge near the phase transition, and write
\begin{equation}
 r_{\rm{YL}} \sim \tau^\Delta |\ln{\tau}|^{\hat{\Delta}}
. \label{edge}
\end{equation}
The gap exponents $\Delta$ and $\hat{\Delta}$ are related to the
more conventional exponents through the relations \cite{Kenna06}
\begin{equation}
 \Delta = \beta + \gamma, \quad \quad \quad \hat{\Delta} = \hat{\beta} - \hat{\gamma} .
\label{gap}
\end{equation}
In \cite{Kenna06}, the Gibbs free energy is written as a function of
$\tau$ and $h$ as
\begin{equation}\label{gibbs}
\Phi(\tau,h) = 2 {\rm{Re}} \int_{r_{\rm{YL}}}^\infty{\ln{\left({ h -
h(r,\tau)  }\right)}} g(r,\tau) dr,
\end{equation}
in which $h(r,\tau)$ is the locus of Lee-Yang zeros in the complex
$h$ plane and where $g(r,\tau)$ is their density. Integrating by
parts yields, for the singular part of the free energy,
\begin{equation}
 \Phi(\tau,h) = -2 {\rm{Re}} \int_{r_{\rm{YL}}}^\infty{\frac{G(r,\tau) \exp{(i\phi)} dr}{h - r \exp{(i\phi)}}}
\,, \label{rain}
\end{equation}
where $G(r,\tau) = \int_{r_{\rm{YL}}(\tau)}^{r}{g(s,\tau) ds}$ is
the cumulative distribution function for the zeros, the locus of
which is assumed to be $h(r,\tau) = r \exp{(i\phi)}$ (the Lee-Yang
theorem gives $\phi = \pi / 2$). In contrast to \cite{Kenna06},
where $h$ was set to zero in (\ref{rain}), the external field is now
kept as a variable here in order to determine its contribution to
scaling near the critical point. From \cite{Kenna06}, the integrated
density is
\begin{equation}
 G (r,\tau) = \chi r_{\rm{YL}}^2 I \left({ \frac{r}{r_{\rm{YL}}} }\right).
\end{equation}
The functional form of $I(x)$ is undetermined here, but our
considerations shall not require such details. Introducing this into
(\ref{rain}), one finds
\begin{equation}
  \Phi(\tau,h) = \chi r_{\rm{YL}}^2 {\cal{F}}_\phi
 \left({ \frac{h}{r_{\rm{YL}}} }\right)
\,,
\end{equation}
where
\begin{equation}
 {\cal{F}}_\phi (y) = -2 {\rm{Re}}\int_1^\infty{\frac{I(x)dx}{y \exp{(-i\phi)}-x}}
\,.
\end{equation}
The specific heat is given by the second derivatives of the free
energy with respect to $\tau$, and is
\begin{equation}
  {\cal{C}}(\tau,h) = \chi r_{\rm{YL}}^2 \tau^{-2} {\cal{F}}_\phi
 \left({ \frac{h}{r_{\rm{YL}}} }\right)
. \label{c}
\end{equation}
Now, from (\ref{rrr013}) and (\ref{edge}), one may express the
scaling of the specific heat in terms of that of the edge:
\begin{equation}
 {\cal{C}}(\tau,h)  =  r_{\rm{YL}}^{2-\frac{\gamma}{\Delta}-\frac{2}{\Delta}} |\ln{r_{\rm{YL}}}|^{\frac{(\gamma+2)\hat{\Delta}}{\Delta}+\hat{\gamma}}  {\cal{F}}_\phi   \left({ \frac{h}{r_{\rm{YL}}} }\right)
, \label{c1}
\end{equation}
which may in turn be written as
\begin{equation}
{\cal{C}}(\tau,h)  =  h^{2-\frac{\gamma}{\Delta}-\frac{2}{\Delta}}
|\ln{h}|^{\frac{(\gamma+2)\hat{\Delta}}{\Delta}+\hat{\gamma}}
{\cal{F}}^\prime_\phi   \left({ \frac{h}{r_{\rm{YL}}} }\right) .
\label{c2}
\end{equation}
Now it is a simple matter to let $\tau \rightarrow 0$ so that
$r_{\rm{YL}} \rightarrow 0$, and the undetermined function
${\cal{F}}^\prime_\phi$ becomes a constant, yielding
\begin{equation}
 {\cal{C}}(h)  = h^{2-\frac{\gamma}{\Delta}-\frac{2}{\Delta}} |\ln{h}|^{\frac{(\gamma+2)\hat{\Delta}}{\Delta}+\hat{\gamma}}  .
\label{c3}
\end{equation}
From the leading behaviour one recovers  (\ref{rrr011}). The
correction exponents lead to the new scaling relation
\begin{equation}
 \hat{\alpha}_c  =  \frac{(\gamma+2) \hat{\Delta}}{\Delta} +\hat{\gamma} 
,
\end{equation}
which, from (\ref{gap}) yields
\begin{equation}\label{newSR1}
\hat{\alpha}_c = \frac{(\gamma + 2)(\hat{\beta}-\hat{\gamma})}
{\beta + \gamma} + \hat{\gamma}.
\end{equation}

Eq.(\ref{rrr010}) and its logarithmic counterpart
\begin{equation}\label{newSR2}
 \hat{\gamma}_c = \hat{\delta} .
\end{equation}
are far more trivial to derive and follow from a single
differentiation of (\ref{rrr015}) with respect to $h$. The latter
two equations (\ref{newSR1}) and (\ref{newSR2}) amount the desired
scaling relations for $\hat{\alpha}_c$ and $\hat{\gamma}_c$.

\section{Thermodynamical functions of a coupled order parameter system on
a scale free network}

In the previous section  we obtained new scaling relations
(\ref{newSR1}), (\ref{newSR2}) for the logarithmic corrections
exponents. Together with the formulas (\ref{rrr018}), (\ref{rrr019})
they form a complete set of scaling relations for the correction to
scaling exponents defined in (\ref{rrr012})--(\ref{rrr017}). The
validity of relations (\ref{rrr018}), (\ref{rrr019}) for spin models
on {\em lattices} was subject to a thorough check in Ref.
\cite{Kenna06}. There, it was shown that the relations hold for all
models where the corrections are known explicitly. In particular,
these include short- and long-range interacting $O(n)$ models at the
upper critical dimension $d_c=4$, spin glasses, percolation and the
Yang-Lee edge problem at $d_c=6$, lattice animals at $d_c=8$,
regular and structurally disordered Ising model at $d=2$, $q$-state
Potts model at $d=2$ and $q_c=4$ (see
\cite{Kenna06,ON_log,ONlr_log,phi3_log,Potts_log,Shalaev94}). Now,
we will proceed further to perform a similar check for the case of
critical behavior on scale-free {\em networks}.

\subsection{Temperature dependencies}

As a case study, we will consider a  rather common situation met in
phase transition theory, when a system exhibits several types of
ordering. This manifests itself by the appearance of two coupled
scalar order parameters denoted by  $x_1$, $x_2$. In a microscopic
description, such a system may be realized as a coupling between two
Ising models, each of them being characterized by its own order
parameter $x_i$, or as an $XY$ model with a single ion anisotropy.
Here we consider the Hamiltonian with a cubic anisotropy term
\begin{equation}\label{cub_ham}
H = -J\sum_{\langle i,j \rangle}\vec{s_i}\cdot\vec{s_j} +
u\sum_{i=1}^{N}\sum_{\nu=1}^{2}s_{\nu,i}^4 ,
\end{equation}
where $\vec{s_i}$ and $\vec{s_j}$ are spins on nodes $i$ and $j$
correspondingly, $J$ and $u$ are the coupling and anisotropy
constants, the index $\nu$ numbers the components of the
two-component vector, $\vec{s_i}\cdot\vec{s_j} = \sum_{\nu=1}^2
s_{\nu,i} s_{\nu,j}$ is a scalar product. The notation
$\sum_{\langle i,j \rangle}$ denotes the summation over all pairs of
connected nodes of the network. Note, that the Hamiltonian
(\ref{cub_ham}) is the free energy of an $n$-vector anisotropic
cubic model in the case $n=2$. The latter is obtained from the
$O(n)$ invariant free energy by adding invariants of the symmetry
group $B_n$ of the $n$-dimensional hypercube \cite{cubicmodel}.

In Ref.~\cite{Palchykov09}, thermodynamical properties of such a
system at $h=0$ were analyzed for a complex scale free network with
a power law node degree distribution exponent $\lambda$, as in
(\ref{rrr001}).  As usual for models on scale-free networks
\cite{spin_on_networks}, the system remains ordered for any finite
temperature for $\lambda \leq 3$, but it possess a second order
phase transition with finite $T_c$ for $\lambda > 3$. The phase
diagram of the system is characterized by two different types of
ordering, either along the edges or along the diagonals of the
square in space the of the order parameter $\vec{x}=\{x_1,x_2\}$. In
the first phase, only one order parameter component is non-zero
($x_1\neq 0$, $x_2=0$ or $x_1= 0$, $x_2 \neq 0$), whereas
$x_1=x_2\neq 0$ in the second phase. The temperature-dependencies of
the order parameter, the susceptibilities and the heat capacities
were obtained and the exponents (\ref{rrr002})--(\ref{rrr004})
determined. The marginal value $\lambda_c=5$ was shown to separate
two different regimes of the phase transition: for $\lambda>5$ the
exponents attain their classical mean field values
\begin{equation}\label{exp1}
\beta=1/2, \hspace{1em} \gamma=1, \hspace{1em} \alpha=0,
\end{equation}
whereas for $3 < \lambda < 5$ two out of the three exponents are
$\lambda$-dependent:
\begin{equation}\label{exp2}
\beta=1/(\lambda-3), \hspace{1em} \gamma=1, \hspace{1em}
\alpha=(\lambda-5)/(\lambda-3).
\end{equation}
Another prominent feature found at $\lambda=5$ for the temperature
dependencies of the order parameter and of the heat capacity is the
appearance of  logarithmic corrections to scaling that have the form
given by Eqs. (\ref{rrr012})--(\ref{rrr014}). The corresponding
correction to scaling exponents were found to be \cite{Palchykov09}
\begin{equation}\label{exp3}
\hat{\beta}=-1/2, \hspace{1em} \hat{\gamma}=0, \hspace{1em}
\hat{\alpha}=-1.
\end{equation}

To complete the analysis of the phase transition in the above model
and to access the leading and correction-to-scaling exponents
((\ref{rrr005})--(\ref{rrr007}) and (\ref{rrr015})--(\ref{rrr017}),
correspondingly) that govern this transition, it is necessary to
analyze the field dependencies of the thermodynamical quantities at
$\tau=0$.

\subsection{General relations}

The starting point of our analysis are the expressions for the free
energy considered for different $\lambda$ in \cite{Palchykov09}
within a Landau-type analysis which was further supported by the
microscopic treatment of the corresponding spin Hamiltonian
(\ref{cub_ham}). In the following, it will be more convenient to
work within the $(T,\vec{x})$-ensemble and to consider the Helmholtz
free energy $F(T,\vec{x})$ related to the Gibbs free energy
$\Phi(T,\vec{h})$, Eq. (\ref{gibbs}), via the Legendre transform
\begin{equation}\label{exp4}
F(T,\vec{x})= \Phi(T,\vec{h}) + \vec{x} \cdot \vec{h}.
\end{equation}
For $\lambda>5$, the free energy reads \cite{Palchykov09}
\begin{equation}\label{feg5}
F(\tau,\vec{x}) = \frac{a}{2}(T-T_c)|\vec{x}|^2 +
\frac{b}{4}|\vec{x}|^4 + \frac{c}{4}x_1^2x_2^2,
\end{equation}
where $|\vec{x}|^2 = x_1^2+x_2^2$. Apart from  the fact that the
parameters $a,b,c$ in (\ref{feg5}) are $\lambda$-dependent (see
\cite{Palchykov09} for explicit expressions), the free energy
(\ref{exp4}) has the form of a usual Landau-type free energy of a
system with coupled scalar order parameters \cite{coupled}.
Therefore, the network structure does not change the critical
exponents for $\lambda>5$.  However, for $\lambda\leq 5$ the leading
terms of the free energy are modified \cite{Palchykov09}:
\begin{eqnarray}\label{fe5}
F(T,\vec{x}) &=& \frac{a}{2}(T-T_c)|\vec{x}|^2 +
\frac{b}{4}|\vec{x}|^4\ln\frac{1}{|\vec{x}|}\\\nonumber
&+&\frac{c}{4}x_1^2x_2^2\ln\frac{1}{|\vec{x}|} \, , \hspace{4em}
\lambda=5 \, , \\ \label{fe35}
 F(T,\vec{x}) &=&
\frac{a}{2}(T-T_c)|\vec{x}|^2 + \frac{b}{4}|\vec{x}|^{\lambda-1} \\
\nonumber &+& \frac{c}{4}x_1^2x_2^2|\vec{x}|^{\lambda-5} \, ,
\hspace{4em} 3<\lambda<5 \, .
\end{eqnarray}
Note that the functional form of the coefficients $a,b,c$ in Eqs.
(\ref{feg5})--(\ref{fe35}) differ, see \cite{Palchykov09} for
detailed formulas. However this explicit form is not important for
 further calculations therefore we keep the same notation for the
coefficients in (\ref{feg5})--(\ref{fe35}).

To analyze the field dependencies of the thermodynamic quantities we
consider that an external magnetic field $h$ is pointing along the
order parameter component $x_1$, $\vec{h}=\{h,0\}$,  and write the
system of equations of state as:
\begin{equation}\label{a1}
\Big(\frac{\partial F(T,\vec{x})}{\partial x_1} \Big)_T = h,
\end{equation}
\begin{equation}\label{a2}
\Big(\frac{\partial F(T,\vec{x})}{\partial x_2} \Big)_T = 0.
\end{equation}
The stable states are determined from the matrix of second
derivatives
\begin{equation}\label{a3}
f_{\mu\eta} = \frac{\partial^2 F(T,\vec{x})}{\partial x_\mu
\partial x_\eta}.
\end{equation}
For a given state, the stability condition requires the real parts
of the eigenvalues of the matrix (\ref{a3}) to be positive. Note
that these eigenvalues are the inverse susceptibilities,
longitudinal $\chi_\parallel^{-1}$ and transverse $\chi_\perp^{-1}$
correspondingly. To derive the heat capacity, one needs to obtain
the entropy of the system. Following the definition
\begin{equation}\label{a5}
S(T,\vec{x}) = -\Big(\frac{\partial F(T,\vec{x})}{\partial T}
\Big)_{\vec{x}}
\end{equation}
one finds the entropy as a function of the temperature $T$ and order
parameter $\vec{x}$. Knowing the dependence of the order parameter
$\vec{x}$ on the temperature and external field $\vec{x} =
\vec{x}(T,\vec{h})$, obtained from the system of equations of state
(\ref{a1}), (\ref{a2}), one finds the entropy as a function of the
temperature and external magnetic field $S = S(T,\vec{h})$. Now the
heat capacity
\begin{equation}\label{a4}
C_h = T\Big(\frac{\partial S(T,\vec{h})}{\partial T}\Big)_h
\end{equation}
completes the calculations. Below, we  sketch the results obtained
for the magnetic field dependencies  of the order parameter,
susceptibilities and of the specific heat for different values of
$\lambda$ at $\tau=0$.

\subsection{Exponents for the dependencies on the magnetic field at $\tau=0$}

Taking the expressions for the free energy (\ref{feg5}) --
(\ref{fe35}) in the equation of state (\ref{a1}) and (\ref{a2}) one
finds the stable solutions. It turns out (see Appendix) that there
are always two stable solutions ($x_1\neq0$, $x_2=0$) and
($x_1\neq0$, $x_2\neq0$). In any case the quantities considered have
the leading form (\ref{rrr005}) -- (\ref{rrr007}). For $\lambda>5$
one obtains the mean field exponents whereas for $3<\lambda<5$ we
arrive at the nontrivial $\lambda$-dependent exponents given in
Table~\ref{tab1}.
\begin{table}[ht]
\begin{center}
\tabcolsep=0.6em
\begin{tabular}{cccccccc}
\hline
             &$\alpha$&$\beta$&$\gamma$&$\delta$&$\alpha_c$&$\gamma_c$ & $\Delta$ \\\hline
  $\lambda\geq5$  &$0$&$1/2$&$1$&$3$&$0$&$2/3$ & 3/2 \\
  $3<\lambda<5$&$\frac{\lambda-5}{\lambda-3}$&$\frac{1}{\lambda-3}$&$1$&$\lambda-2$&$\frac{\lambda-5}{\lambda-2}$&$\frac{\lambda-3}{\lambda-2}$ &
  $\frac{\lambda-2}{\lambda-3}$ \\
\hline
\end{tabular}
\end{center}
\caption{\label{tab1} Critical exponents governing temperature and
field dependencies of thermodynamic quantities for different values
of $\lambda$.}
\end{table}

At the marginal value $\lambda=5$ the logarithmic corrections of the
form (\ref{rrr015}) -- (\ref{rrr017}) are obtained as summarized in
Table~\ref{tab2} completed by the gap exponents calculated via
Eq.~(\ref{gap}).
\begin{table*}[ht]
\begin{center}
\tabcolsep=0.7em
\begin{tabular}{l|ccccccc}
\hline
 &$\hat{\alpha}$&$\hat{\beta}$&$\hat{\gamma}$&$\hat{\delta}$&$\hat{\alpha_c}$&$\hat{\gamma_c}$ & $\hat{\Delta}$\\\hline
 scale-free network, $\lambda=5$ &$-1$&$-1/2$&$0$&$-1/3$&$-1$&$-1/3$ & $-1/2$ \\\hline
$q=4$ Potts, $d=2$ &$-1$&$-1/8$&$3/4$&$-1/15$&$-22/15$&$-1/15$ &
$-7/8$
\\\hline
$d=4$ $O(n)$ model
&$\frac{4-n}{n+8}$&$\frac{3}{n+8}$&$\frac{n+2}{n+8}$&$1/3$&$-$&$-$ &
$\frac{1-n}{n+8}$
\\\hline
\end{tabular}
\end{center}
\caption{\label{tab2} Exponents for the logarithmic corrections to
scaling laws that appear for several models: spin model on a scale
free network at marginal value $\lambda=5$ (our results); $d=2$
Potts model at marginal spin states number $q=4$; and $O(n)$
symmetric model at marginal space dimension $d=4$ (see
\cite{Kenna06} and references therein, \cite{Shchur09}).}
\end{table*}

With the data of Table \ref{tab2} at hand, it is straightforward to
check the validity of the scaling relations for the logarithmic
correction to scaling exponents (\ref{rrr018}), (\ref{rrr019}),
(\ref{newSR1}), (\ref{newSR2}). Moreover, one can see that they
constitute a separate family that differs from other logarithmic
correction-to-scaling exponents. To this end, we give in the
Table~\ref{tab2} the value of the logarithmic correction-to-scaling
exponents that arise for $d=2$ Potts model at marginal number of
spin states $q=4$ and for the $O(n)$-symmetrical model at marginal
space dimension $d=4$. Table~\ref{tab1} also gives further evidence
of the validity of scaling relations (\ref{rrr008})--(\ref{rrr011})
for the leading scaling exponents in particular of those that
involve $\alpha_c$ and $\gamma_c$. Values of the latter exponents
for scale-free networks have not been available before.

\section{Scaling functions}

Finally, we consider how the underlying structure in the form of a
complex network affects the validity of the scaling hypothesis, and
if the latter is satisfied, we will find corresponding scaling
functions. The hypothesis states that the singular part of a
thermodynamic potential of a system in the vicinity of the critical
point has the form of a generalized homogeneous function
\cite{Stanley_Domb}. For the Helmholtz potential this statement can
be mathematically written as \cite{note1}
\begin{equation}\label{eqq02}
F(\tau,m) = \tau^{2-\alpha} f_{\pm}(m/\tau^{\beta}),
\end{equation}
where the sign $\pm$ corresponds to $T>T_c$ or $T<T_c$ respectively.
Provided the homogeneity hypothesis for the potential (\ref{eqq02})
holds, one arrives at the scaling form for the other thermodynamic
quantities \cite{comment_Hankey}. In particular, below we will make
use of three different equivalent representations for the equation
of state \cite{Hankey72,scaling_functions}. The Widom-Griffiths
scaling form of the equation of state is \cite{scaling_functions}:
\begin{equation}\label{eqq03}
h = m^\delta h_{\pm}(\tau/m^{1/\beta}),
\end{equation}
with the alternative representation
\begin{equation}\label{eqq04}
h = \tau^{\beta\delta} H_{\pm}(m/\tau^{\beta}).
\end{equation}
The scaling form of the magnetization reads (see also
\cite{comment_Hankey}):
\begin{equation}\label{eqq05}
m = \tau^\beta \mu_{\pm}(h/\tau^{\beta\delta}),
\end{equation}
and the isothermal susceptibility may be written as
\begin{equation}\label{eqq06}
\chi_T = \tau^{-\gamma} \chi_{\pm}(h/\tau^{\beta\delta}).
\end{equation}
Note that taking temperature derivatives of (\ref{eqq02}) one
arrives at the scaling functions for the entropy and heat capacity.
Since their derivation follows in a similar manner as the above
introduced functions (\ref{eqq03}) -- (\ref{eqq06}) we do not give
their explicit expressions here.

The formulae given above hold for the single scalar order parameter
system, from which we will start our consideration. Then the system
of $O(n)$ symmetrical vector order parameter $\vec{m}=\{m_1,m_2\}$
and the system of coupled order parameters will be analyzed, for
which one may easily generalize Eqs.(\ref{eqq02})--(\ref{eqq06}).

\subsection{Single order parameter}

For $\lambda>5$ the Helmholtz potential $F(\tau,m)$ for the system
with a single order parameter (magnetization) $m$ may be obtained
from Eq.(\ref{feg5}) substituting $x_1=m$, $x_2=0$. Then $F(\tau,m)$
in dimensionless variables is
\begin{equation}\label{eqq07}
F(\tau,m) = \pm\frac{1}{2}\tau m^2 + \frac{1}{4}m^4,
\end{equation}
where the energy is measured in units of $F_0$, and the
magnetization in units of $m_0$:
\begin{equation}\label{eqq09}
F_0 = a^2T_c^2/b, \hspace{1em} m_0 =
\sqrt{\frac{aT_c}{b}}.
\end{equation}
It is easy to see that $F(\tau,m)$ scales as
\begin{equation}\label{eqq09a}
F(\tau,m) = \tau^2 f_{\pm}\big(m/\tau^{1/2}\big),
\end{equation}
where
\begin{equation}\label{eqq09b}
f_{\pm}(\zeta) = \pm\frac{1}{2}\zeta^2 + \frac{1}{4}\zeta^4.
\end{equation}

For $\lambda=5$ due to the logarithmic corrections in the free
energy (\ref{fe5}), the scaling form defined above fails. We refrain
from giving a scaling function for this case.

For $3<\lambda<5$, $F(\tau,m)$ as given by Eq.(\ref{fe35}) for the
single order parameter system reads
\begin{equation}\label{eqq28}
F(\tau,m) = \pm\frac{1}{2}\tau m^2 + \frac{1}{4}m^{\lambda-1}.
\end{equation}
Now the dimensionful quantities $F_0$ and $m_0$ (\ref{eqq09}) become
$\lambda$-dependent:
\begin{equation}\label{eq30}
F_0(\lambda) =
\frac{(aT_c)^{\frac{\lambda-1}{\lambda-3}}}{b^{1/(\lambda-3)}},\hspace{1em}
m_0(\lambda) = \Big(\frac{aT_c}{b}\Big)^{\frac{1}{\lambda-3}}.
\end{equation}

Again, one can recast (\ref{eqq28}) singling out the scaling
function as:
\begin{equation}\label{eq30a}
F(\tau,m) = \tau^{\frac{\lambda-1}{\lambda-3}}
f_{\pm}\big(m/\tau^{1/(\lambda-3)}\big),
\end{equation}
where the scaling function $f_{\pm}(\zeta)$ acquires a
$\lambda$-dependence,
\begin{equation}\label{eq30b}
f_{\pm}(\zeta) = \pm\frac{1}{2}\zeta^2 +
\frac{1}{4}\zeta^{\lambda-1},
\end{equation}
and we have taken into account the $\lambda$-dependence of the heat
capacity critical exponent $\alpha=(\lambda-5)/(\lambda-3)$, see
Table~\ref{tab1}.

\begin{figure*}
\parbox{8cm}{\includegraphics[width=8cm]{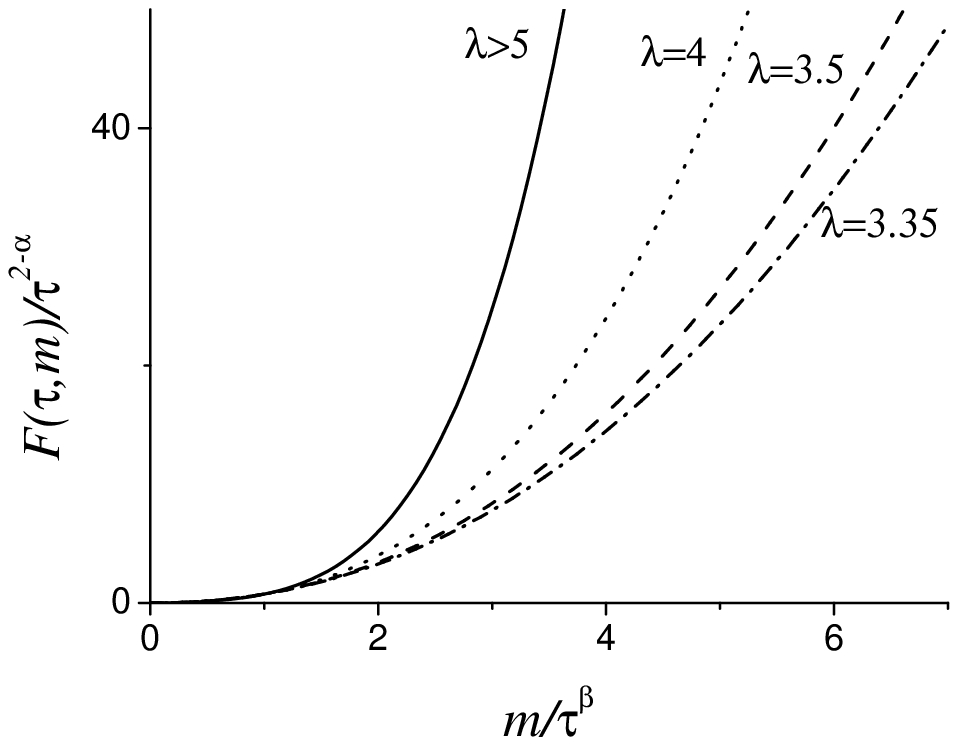}}
\parbox{8cm}{\includegraphics[width=8cm]{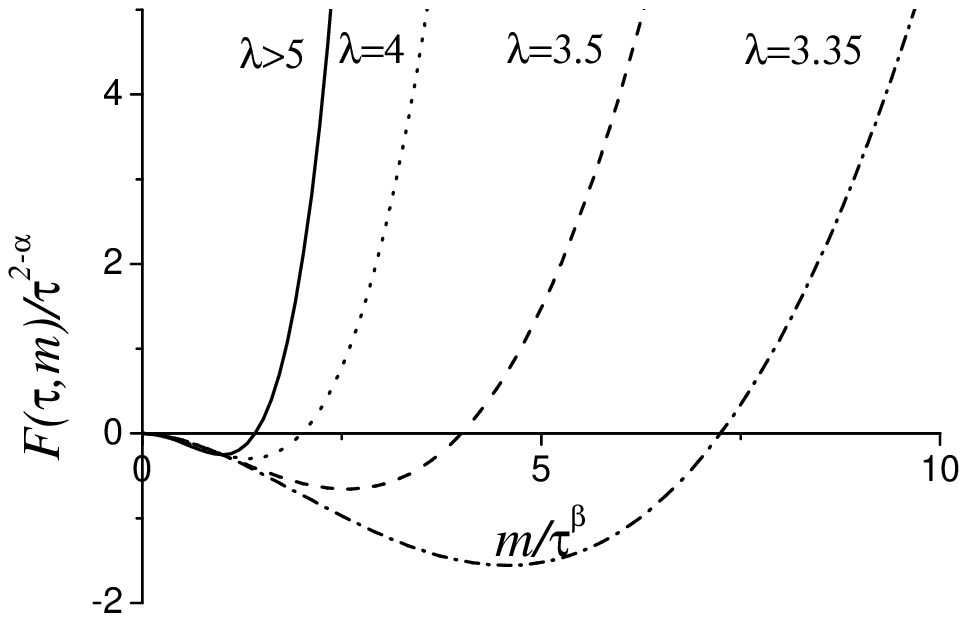}}\\
\parbox{8cm}{\bf (a)}
\parbox{8cm}{\bf (b)}
 \caption{The dependency of $F(\tau,m)/\tau^{2-\alpha}$ on
$m/\tau^{\beta}$ for values of $\lambda$ between 5 and 3. Plot ({\bf
a}) applies to temperatures above $T>T_c$ while ({\bf b}) applies to
$T<T_c0$.}
 \label{fig1}
\end{figure*}

In Fig.~\ref{fig1} we show the dependence of
$F(\tau,m)/\tau^{2-\alpha}$ on $m/\tau^{\beta}$ for different values
of $\lambda$ above and below the critical temperature. Here and
below we will plot the corresponding scaling function in the region
of positive values of $h$ and $m$.

\begin{figure}
\centerline{
\includegraphics[width=80mm]{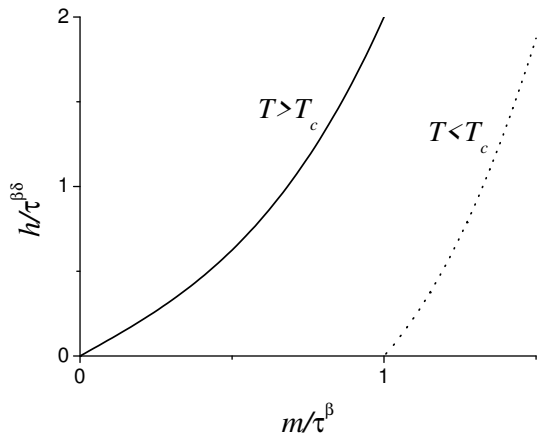}}
\caption{The dependence of $h/\tau^{\beta\delta}$ on
$m/\tau^{\beta}$ given for $\lambda>5$ by the scaling function
$H_{\pm}$ (\ref{eqqq02}), above (solid line) and below (dotted line)
the critical temperature.}
 \label{fig2}
\end{figure}

\begin{figure}
\centerline{
\includegraphics[width=80mm]{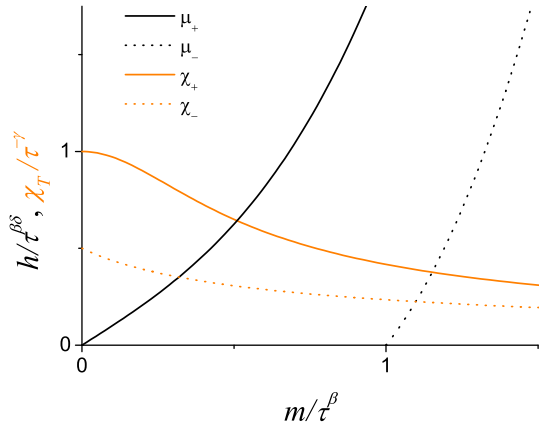}}
\caption{The scaling functions for magnetization and isothermal
magnetic susceptibility for $\lambda>5$ above (solid lines) and
below (dotted lines) the critical temperature. Black curves:
$m/\tau^{\beta}$, light (orange) curves: isothermal susceptibility
scaling function $\chi_{\pm}$ (color online).}\label{fig3}
\end{figure}

Now let us consider the scaling functions $h_{\pm}$, $H_{\pm}$,
$\mu_{\pm}$ and $\chi_{\pm}$ defined by
Eqs.(\ref{eqq03})--(\ref{eqq06}). For the case $\lambda>5$ one finds
from (\ref{eqq07}) the equation of state
\begin{equation}\label{eqqq03}
m^3 \pm \tau m - h=0.
\end{equation}
Representing (\ref{eqqq03}) as defined by (\ref{eqq03}) we arrive at
the scaling function $h_{\pm}(\zeta)$ that describes the equation of
state in the Widom-Griffiths form
\begin{equation}\label{eqqq01}
h_{\pm}(\zeta) = 1\pm \zeta,
\end{equation}
and the scaling function, defined by Eq.(\ref{eqq04}) readily
follows
\begin{equation}\label{eqqq02}
H_{\pm}(\zeta) = \zeta^3 \pm \zeta .
\end{equation}
The dependence of $h/\tau^{\beta\delta}$ on $m/\tau^{\beta}$, given
by the scaling function $H_{\pm}$ in Eq.(\ref{eqqq02}) is shown in
Fig.~\ref{fig2}.
To get the magnetization scaling function $\mu_{\pm}$, given by
Eq.~(\ref{eqq05}), we first note that $\mu_{\pm}$ express the
dependence of $m/\tau^{\beta}$ on $h/\tau^{\beta\delta}$, which is
inverse to the dependence, given by the function $H_{\pm}$ in
Eq.~(\ref{eqq04}). Correspondingly, the scaling function $\mu_{\pm}$
may be easily plotted by exchanging the axes in Fig.~\ref{fig2} as
shown in Fig.~\ref{fig3} by black bold lines. Three analytic
solutions of the equation of state (\ref{eqqq03}) for $m$ give three
branches for the function $\mu_{\pm}$. Above the critical
temperature only one branch is real and is presented by the black
solid line in Fig.~\ref{fig3}. The black dotted line in
Fig.~\ref{fig3} displays the scaling function $\mu_{-}$ below the
critical temperature $T<T_c$, which corresponds to the real solution
of Eq.~(\ref{eqqq03}). The above described solutions are given by
the following formulae
\begin{equation}\label{eq109}
\mu_{\pm}(\zeta) = \frac{\varphi_{\pm}(\zeta)}{6} \mp
\frac{2}{\varphi_{\pm}(\zeta)},
\end{equation}
where
\begin{equation}
\varphi_{\pm}(\zeta) = \Big(108\zeta + 12\sqrt{81\zeta^2\pm
12}\Big)^{1/3}.
\end{equation}

The scaling function for the susceptibility (\ref{eqq06}) may be
easily obtained from (\ref{eq109}) using the relation
\begin{equation}\label{eqq15}
\chi_{\pm}(\zeta) = \frac{{\rm d} \mu_{\pm}(\zeta)}{{\rm d} \zeta}.
\end{equation}
and it reads
\begin{equation}\label{eqq15a}
\chi_{\pm}(\zeta) = \Big(\frac{\varphi_{\pm}(\zeta)}{6} \pm
\frac{2}{\varphi_{\pm}(\zeta)}\Big)\frac{36}{\varphi_{\pm}^3(\zeta)}\Big(1+\frac{9\zeta}{\sqrt{81\zeta^2\pm12}}\Big).
\end{equation}
The dependence of $\chi_T/\tau^{-\gamma}$ on $h/\tau^{\beta\delta}$,
described by the scaling function (\ref{eqq15}) is plotted in
Fig.~\ref{fig3} by light (orange online) lines.

As we have observed above, the scaling hypothesis for the Helmholtz
free energy holds for $\lambda>5$ and $3<\lambda<5$. To proceed
further and to get the scaling functions $h_{\pm}$, $H_{\pm}$,
$\mu_{\pm}$ and $\chi_{\pm}$ in the region $3<\lambda<5$ we first
derive from (\ref{eqq28}) an equation of state, which now has the
form
\begin{equation}\label{z1}
\frac{\lambda-1}{4}m^{\lambda-2}\pm\tau m-h=0.
\end{equation}
Again, expressing $h$ in terms of $m$ and making use of
Eq.~(\ref{eqq03}) we get for the scaling function $h_{\pm}$ that
enters the equation of state in Widom-Griffiths form,
\begin{equation}\label{eqqa01}
h_{\pm}(\zeta) = \frac{\lambda-1}{4}\pm\zeta,
\end{equation}
Note, that in the region $3<\lambda<5$ some of the critical indices
acquire $\lambda$ dependence. In particular, to get (\ref{eqqa01})
one should take into account that $\beta=1/(\lambda-3)$ and
$\delta=\lambda-2$, (Table~\ref{tab1}). Comparing
Eqs.~(\ref{eqqa01}) and (\ref{eqqq01}) one can see, that for
$3<\lambda<5$ the functional dependence of $h_{\pm}$ on $\zeta$ does
not change, and the only difference is the $\lambda$-dependence of
the first term in the right-hand side of Eq.~(\ref{eqqa01}). This is
not the case for the function $H_{\pm}$. Indeed, from (\ref{z1}) and
(\ref{eqq04}) we get for this function
\begin{equation}\label{eqqa02}
H_{\pm}(\zeta) = \frac{\lambda-1}{4}\zeta^{\lambda-2} \pm \zeta.
\end{equation}
Now not only the coefficient but also the leading power of this
function is $\lambda$-dependent.

\begin{figure}
\parbox{8.5cm}{\includegraphics[width=8.5cm]{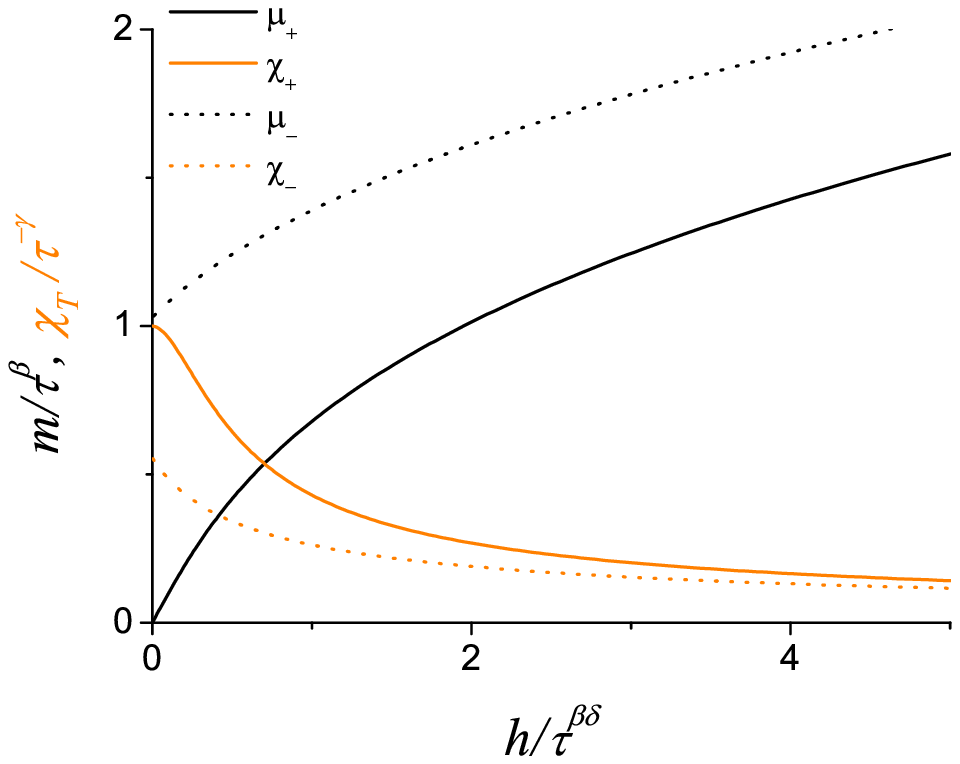}}\\
\parbox{8cm}{\bf (a)}\\
\parbox{8.5cm}{\includegraphics[width=8.5cm]{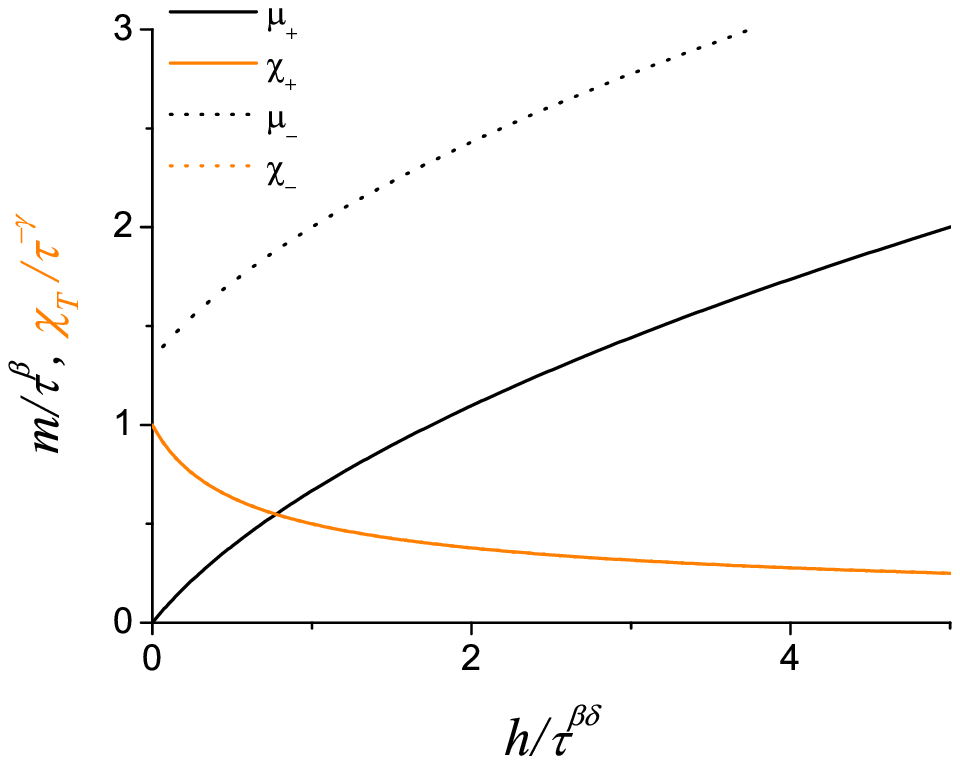}}\\
\parbox{8cm}{\bf (b)}\\
\parbox{8.5cm}{\includegraphics[width=8.5cm]{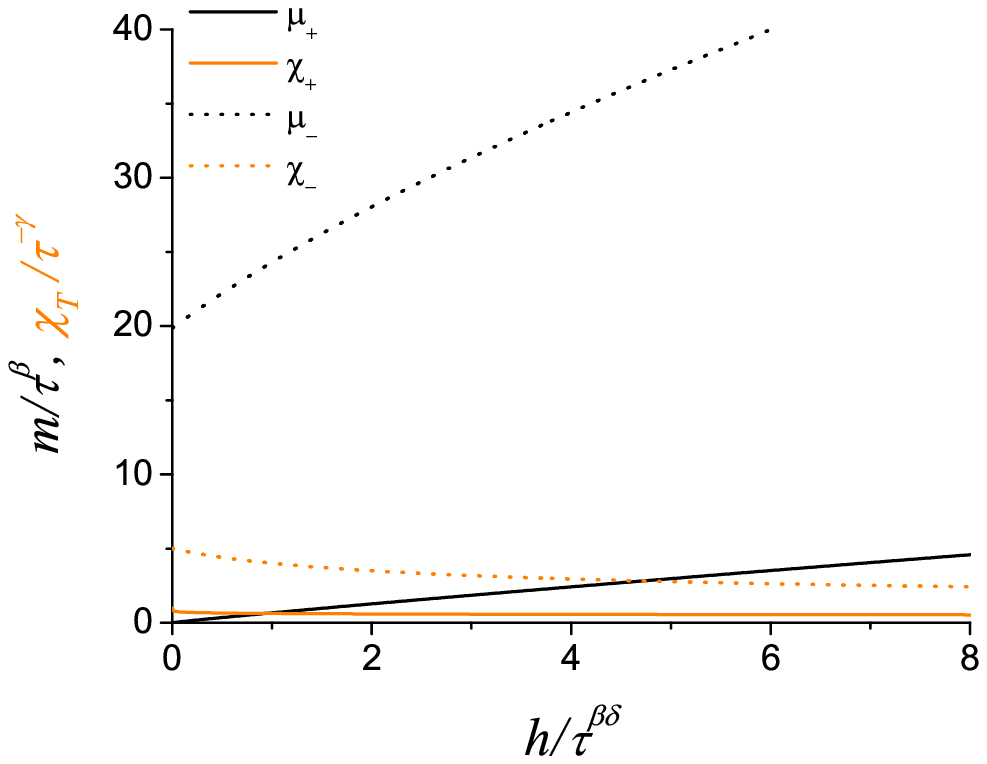}}\\
\parbox{8cm}{\bf (c)}
 \caption{The scaling functions for magnetization and isothermal
magnetic susceptibility for $\lambda=4.8$ ({\bf a}), $\lambda=4.0$
({\bf b}), $\lambda=3.1$ ({\bf c}) above (solid lines) and below
(dotted lines) the critical temperature. Black curves:
$m/\tau^{\beta}$, light (orange) curves: isothermal susceptibility
scaling function $\chi_{\pm}$.}\label{fig4}
\end{figure}

There is one more observation which follows from the comparison of
Eqs.~(\ref{eqqa02}) and (\ref{eqqq02}) that express function
$H_{\pm}$ for $3<\lambda<5$ and $\lambda>5$, correspondingly.
Eq.(\ref{eqqq02}) allows for an analytic solution, which enables us
in particular to find an analytic form for the scaling functions
$\mu_{\pm}$, $\chi_{\pm}$ at $\lambda>5$ (see Eq.(\ref{eq109}) and
(\ref{eqq15a})). Whereas a similar analytic treatment is possible
for $H_{\pm}$ at integer values of the power ($\lambda-2$) (i.e. for
$\lambda=4$), it is impossible for the general non-integer value of
$3<\lambda<5$. Therefore, we make use of the graphic representation
to show the behavior of $\mu{\pm}$, $\chi_{\pm}$ at different
$\lambda$ in Fig.~\ref{fig4}.

Comparing the plots in Fig.~\ref{fig4} one observes a particular
feature in the behavior of the isothermal susceptibility scaling
functions $\chi_{\pm}$. In the region $4<\lambda<5$ the curve
$\chi_{+}$ is above the corresponding curve $\chi_{-}$ for all
values of the argument. Qualitatively this resembles the case
$\lambda>5$, where the usual Landau theory (\ref{eqq07}) holds. In
particular, such behavior means that the plot for the zero-field
isothermal susceptibility $\chi(\tau)$ has the usual
'$\lambda$-shape' in the vicinity of $\tau=0$, i.e. the left
shoulder of the curve is lower than the right one. This situation
changes with a further decrease of $\lambda$: first at $\lambda=4$
both curves $\chi_{+}$ and $\chi_{-}$ coincide, and then, for
$3<\lambda<4$ the curve $\chi_{+}$ is below $\chi_{-}$. Again, for
the zero-field isothermal magnetic susceptibility $\chi(\tau)$ this
would mean that its left shoulder ($T<T_c$) is above its right one
($T>T_c$): the usual '$\lambda$-shape' turns to a
'mirror-inversed-$\lambda$-shape'. The latter is closely related to
the universal amplitude ratios of the magnetic susceptibility
$\Gamma_{+}/\Gamma_{-}=\lambda-3$, considered in
Ref.~\cite{Palchykov09}.

\subsection{Vector order parameter}

Now let us consider a system with an $O(n)$ symmetric vector order
parameter being in particular interested in the $n=2$ case, when
$\vec{m}=\{m_1,m_2\}$. The free energy $F(\tau,\vec{m})$ for such a
system may be obtained from Eqs.~(\ref{feg5})--(\ref{fe35}) by
excluding the coupling term. So, for $\lambda>5$, $F(\tau,\vec{m})$
in the dimensionless variables is
\begin{equation}\label{eqv1}
F(\tau,m) = \pm\frac{1}{2}\tau |\vec{m}|^2 + \frac{1}{4}|\vec{m}|^4,
\end{equation}
where the units of measure are given by Eq.~(\ref{eqq09}). For the
vector order parameter the free energy scaling function
$f_{\pm}(\vec{\zeta})$ is defined by the equation
\begin{equation}\label{fsc}
F(\tau,\vec{m}) = \tau^{2-\alpha}
f_{\pm}\big(\vec{m}/\tau^{\beta}\big),
\end{equation}
and reads
\begin{equation}
f_{\pm}(\vec{\zeta}) = \pm\frac{1}{2}|\vec{\zeta}|^2 +
\frac{1}{4}|\vec{\zeta}|^4.
\end{equation}
Taking that the magnetic field points along the $m_1$ component, one
finds that the stable state of the system requires
\begin{equation}\label{eqs1}
m_1^3 \pm \tau m_1 - h=0,
\end{equation}
and
\begin{equation}\label{eqs2}
m_2=0,
\end{equation}
as follows from the system of equations of state (\ref{a1}),
(\ref{a2}). Since the field points only along the first component of
the magnetization, we get that $\mu_{2\pm}=0$ and
$\mu_{1\pm}\equiv\mu_{\pm}$. The scaling functions $h_{\pm}$,
$H_{\pm}$ and $\mu_{1\pm}\equiv\mu_{\pm}$ coincide with the
corresponding scaling functions for the single scalar order
parameter, and are given by Eqs.~(\ref{eqqq01})--(\ref{eq109}) where
$m$ is to be replaced by $m_1$. Nevertheless, one needs to consider
two different response functions, that describe the reaction of the
system on an external magnetic field: the longitudinal and
transverse susceptibilities, which scale as
\begin{equation}\label{eqc22a}
\chi_{\parallel} = \tau^{-\gamma}
\chi_{\parallel\pm}(h/\tau^{\beta\delta}),
\end{equation}
\begin{equation}\label{eqc22b}
\chi_{\perp} = \tau^{-\gamma} \chi_{\perp\pm}(h/\tau^{\beta\delta}).
\end{equation}
Using the definition for $\chi_{\parallel}$, $\chi_{\perp}$ (see
Eq.~(\ref{a3}) and below) and substituting solutions (\ref{eqs1}),
(\ref{eqs2}) into corresponding derivatives of the free energy
(\ref{eqv1}) we arrive at the scaling functions
\begin{equation}
\chi_{\parallel\pm}(\zeta) = \Big[ \pm1 +
3\big(\mu_{\pm}(\zeta)\big)^2\Big]^{-1},
\end{equation}
\begin{equation}
\chi_{\perp\pm}(\zeta) = \Big[ \pm1 +
\big(\mu_{\pm}(\zeta)\big)^2\Big]^{-1},
\end{equation}
with $\mu_{\pm}(\zeta)$ given by Eq.~(\ref{eq109}). Note that the
existence of the continuous $O(n)$ symmetry of the free energy at
$h=0$ allows also to present the transverse susceptibility as a
ratio of $m_1$ and $h$ at arbitrary $h$:
\begin{equation}\label{chi_p}
\chi_{\perp} = \frac{m_1}{h}.
\end{equation}

Let us repeat the above calculations for $3<\lambda<5$. Excluding
the coupling term from Eq.~(\ref{fe35}) we get for the $O(n)$
symmetric free energy:
\begin{equation}
F(\tau,\vec{m}) = \pm\frac{1}{2}\tau |\vec{m}|^2 +
\frac{1}{4}|\vec{m}|^{\lambda-1},
\end{equation}
and the scaling function $f_{\pm}$ follows
\begin{equation}
f_{\pm}(\vec{\zeta}) = \pm\frac{1}{2}|\vec{\zeta}|^2 +
\frac{1}{4}|\vec{\zeta}|^{\lambda-1}.
\end{equation}
The units of measure are given by Eq.~(\ref{eq30}). Again, the
system of equations of state (\ref{a1}), (\ref{a2}) requires
\begin{equation}
m_2=0,
\end{equation}
and the equation for $m_1$ coincides with Eq.~(\ref{z1}) for a
scalar order parameter $m$:
\begin{equation}
\frac{\lambda-1}{4}m_1^{\lambda-2} \pm \tau m_1 - h=0.
\end{equation}
Properly, the scaling functions $h_{\pm}$, $H_{\pm}$ and
$\mu_{1\pm}\equiv\mu_{\pm}$ are equal to the corresponding scaling
functions for the single scalar order parameter. The
susceptibilities scaling functions $\chi_{\parallel\pm}$ and
$\chi_{\perp\pm}$ follow
\begin{equation}
\chi_{\parallel\pm}(\zeta) = \Big[ \pm1 +
\frac{(\lambda-1)(\lambda-2)}{4}\big(\mu_{1\pm}(\zeta)\big)^{\lambda-3}\Big]^{-1},
\end{equation}
\begin{equation}
\chi_{\perp\pm}(\zeta) = \Big[ \pm1 +
\frac{\lambda-1}{4}\big(\mu_{1\pm}(\zeta)\big)^{\lambda-3}\Big]^{-1}.
\end{equation}

Again, as for the case $\lambda>5$ we note that the transverse
susceptibility $\chi_{\perp}$ also for $3<\lambda<5$ can be recast
as the Eq.~(\ref{chi_p}).

\subsection{Coupled order parameters}

For the model we consider in this paper, the vector $\vec{m}$ has
two components which corresponds to two coupled scalar order
parameters $m_1$ and $m_2$. For the case $\lambda>5$ the free energy
is given by Eq.~(\ref{feg5}) and the corresponding scaling function
follows
\begin{equation}
f_{\pm}(\vec{\zeta}) = \pm\frac{1}{2}|\vec{\zeta}|^2 +
\frac{1}{4}|\vec{\zeta}|^4 + \frac{c}{4b}\zeta_1^2\zeta_2^2,
\end{equation}
where the units of measure are given by Eq.~(\ref{eqq09}).

Let us recall, that the presence of the coupling between the order
parameters $m_1$ and $m_2$ leads to two possible stable states of
the system. Besides the state
\begin{equation}\label{m10}
[m_1,0],\hspace{1em} {\rm with}\hspace{1ex} m_1\neq0, m_2=0,
\end{equation}
the system is characterized by an additional stable state:
\begin{equation}\label{m1m2}
[m_1,m_2],\hspace{1ex} {\rm with}\hspace{1ex} m_1\neq0, m_2\neq0.
\end{equation}

In the ordered state $[m_1,0]$, where only $m_1$ reflects the system
magnetization, the scaling functions $h_{\pm}$, $H_{\pm}$ and
$\mu_{1\pm}\equiv\mu_{\pm}$ coincide with the single scalar order
parameter given in Eqs.~(\ref{eqqq01})--(\ref{eq109}). For the type
of ordering considered, $[m_1,0]$, the susceptibility scaling
functions read:
\begin{equation}\label{eq40w}
\chi_{\parallel\pm}(\zeta) = \Big[ \pm1 +
3\big(\mu_{\pm}(\zeta)\big)^2\Big]^{-1},
\end{equation}
\begin{equation}\label{eq40w1}
\chi_{\perp\pm}(\zeta) = \Big[ \pm1 +
\frac{2b+c}{2b}\big(\mu_{\pm}(\zeta)\big)^2\Big]^{-1},
\end{equation}
with $\mu_{\pm}(\zeta)$ given by Eq.~(\ref{eq109}).

For the second type of ordering, $[m_1,m_2]$, the system of
equations of state (\ref{eq49}), (\ref{eq50}) may be written as
\begin{eqnarray}\label{eqcc16}
\left\{\begin{array}{ccl}
m_1^3 \pm \tilde{\tau} m_1 - \tilde{h}=0,\\
m_2^2 = -\frac{2b+c}{2b}m_1^2 -\tau,
\end{array}\right.
\end{eqnarray}
where
\begin{equation}\label{eqq92}
\tilde{\tau} = \frac{2b}{4b+c}\tau,\hspace{2em} \tilde{h} =
-\frac{4b^2}{c(4b+c)}h.
\end{equation}
The coefficient in front of $\tau$ in Eq.~(\ref{eqq92}) is positive
due to the stability condition in the vicinity of the critical point
at zero magnetic field ($h=0$) \cite{Palchykov09} as well as at the
critical temperature ($\tau=0$) with applied field ($h\neq0$),
Eq.~(\ref{s2}).

The Widom-Griffiths form of the equation of state (\ref{eqq03})
reads
\begin{equation}\label{eq011b}
\tilde{h} = m_1^3 h_{\pm}(\tilde{\tau}/m_1^{1/2}),
\end{equation}
with $h_{\pm}(\zeta)$, given by Eq.~(\ref{eqqq01}).
The equation of state, given by (\ref{eqq04}) is then
\begin{equation}\label{eq021b}
\tilde{h} = \tilde{\tau}^{3/2} H_{\pm}(m_1/\tilde{\tau}^{1/2}),
\end{equation}
with $H_{\pm}(\zeta)$ given by Eq.(\ref{eqqq02}).

The order parameters $m_1$ and $m_2$, obtained from the system
(\ref{eqcc16}) may be conveniently presented in the scaling form
\begin{eqnarray}\label{eqcc17}
m_1 &=& \tilde{\tau}^{1/2}\mu_{1\pm}\Big(
\tilde{h}/\tilde{\tau}^{3/2}\Big),\\\nonumber m_2 &=&
\tilde{\tau}^{1/2}\mu_{2\pm}\Big( \tilde{h}/\tilde{\tau}^{3/2}\Big).
\end{eqnarray}
The scaling function $\mu_{1\pm}(\zeta)$ coincides with the function
$\mu_{\pm}(\zeta)$ in (\ref{eq109}) of the single order parameter
system, and for $\mu_{2\pm}(\zeta)$ we find:
\begin{equation}
\mu_{2\pm}^2(\zeta) = -\frac{2b+c}{2b}\mu_{1\pm}^2(\zeta)
\mp\frac{4b+c}{2b}.
\end{equation}

The susceptibilities follow the scaling form (\ref{eqc22a}),
(\ref{eqc22b}) where $\chi_{\parallel\pm}(\zeta)$ and
$\chi_{\perp\pm}(\zeta)$ are
\begin{widetext}
\begin{equation}\label{eq40wa}
\chi_{\parallel\pm}(\zeta) = \Big[ \pm1 +
\frac{8b+c}{4b}\big(\mu_1^2 + \mu_2^2\big)
+\frac{1}{4}\sqrt{\big(4-\frac{c}{b}\big)^2\big(\mu_1^4 +
\mu_2^4\big) + 2\big(16 + 40\frac{c}{b} +
7\frac{c^2}{b^2}\big)\mu_1^2\mu_2^2} \Big]^{-1}.
\end{equation}
\begin{equation}\label{eq40wb}
\chi_{\perp\pm}(\zeta) = \Big[ \pm1 + \frac{8b+c}{4b}\big(\mu_1^2 +
\mu_2^2\big) -
\frac{1}{4}\sqrt{\big(4-\frac{c}{b}\big)^2\big(\mu_1^4 +
\mu_2^4\big) + 2\big(16 + 40\frac{c}{b} +
7\frac{c^2}{b^2}\big)\mu_1^2\mu_2^2} \Big]^{-1}.
\end{equation}
\end{widetext}
Here we used the notations $\mu_1 \equiv \mu_{1\pm}(\zeta)$, $\mu_2
\equiv \mu_{2\pm}(\zeta)$.

For the case $3<\lambda<5$ the Helmholtz potential (\ref{fe35})
follows the scaling form (\ref{fsc}) where the scaling function
\begin{equation}\label{eqcd13b}
f_{\pm}(\vec{\zeta}) = \pm\frac{1}{2}|\vec{\zeta}|^2 +
\frac{1}{4}|\vec{\zeta}|^{\lambda-1} +
\frac{c}{4b}\zeta_1^2\zeta_2^2|\vec{\zeta}|^{\lambda-5}
\end{equation}
becomes functionally $\lambda$-dependent.

Similarly as for $\lambda>5$, for $3<\lambda<5$ the system of
equations of state (\ref{a20}), (\ref{a21}) permits two types of
ordering at nonzero external field $h\neq0$: $[m_1,0]$
(Eq.~(\ref{m10})) and $[m_1, m_2]$ (Eq.~(\ref{m1m2})).

For the case $[m_1,0]$ where only $m_1$ depends on the external
field, the scaling functions $h_{\pm}$ and $H_{\pm}$ coincide with
the scaling functions given by Eqs.~(\ref{eqqa01}), (\ref{eqqa02}).
Similarly, as for the single order parameter, the analytic
treatments for the magnetization and the isothermal susceptibilities
are impossible for general non-integer $\lambda$, for which
$3<\lambda<5$, whereas their shape for different $\lambda$ is shown
in Fig.~\ref{fig4}. Note however, that the isothermal susceptibility
scaling functions, defined by Eqs.~(\ref{eqc22a}), (\ref{eqc22b})
may be analytically expressed through the magnetization scaling
function $\mu_{1\pm}$ as
\begin{equation}
\chi_{\parallel\pm}(\zeta) = \Big( \pm1 +
\frac{(\lambda-1)(\lambda-2)}{4}\big[\mu_{1\pm}(\zeta)\big]^{\lambda-3}\Big)^{-1},
\end{equation}
\begin{equation}
\chi_{\perp\pm}(\zeta) = \Big( \pm1 + \Big[\frac{\lambda-1}{4} +
\frac{c}{2b}\Big]\big[\mu_{1\pm}(\zeta)\big]^{\lambda-3}\Big)^{-1}.
\end{equation}

For the ordering $[m_1,m_2]$, the equations of state (\ref{a20}),
(\ref{a21}) do not allow to obtain analytical solutions for $m_1$
and $m_2$ in the general case for arbitrary non-integer $\lambda$.
Nevertheless, these equations enable a confirmation of the scaling
of the magnetization. Indeed, substituting
\begin{equation}
m_1 = \tau^{1/(\lambda-3)}
\mu_{1\pm}\big(h/\tau^{(\lambda-2)/(\lambda-3)}\big),
\end{equation}
\begin{equation}
m_2 = \tau^{1/(\lambda-3)}
\mu_{2\pm}\big(h/\tau^{(\lambda-2)/(\lambda-3)}\big)
\end{equation}
into Eqs.~(\ref{a20}), (\ref{a21}) one obtains the following
equation for the scaling functions $\mu_{1\pm}$ and $\mu_{2\pm}$:
\begin{equation}\label{e111}
\frac{c}{2b}\mu_{1\pm}(\zeta)\Big(\mu_{2\pm}^2(\zeta)-\mu_{1\pm}^2(\zeta)\Big)|\vec{\mu}_{\pm}(\zeta)|^{\lambda-5}=\zeta,
\end{equation}
where the relation between $\mu_{1\pm}(\zeta)\equiv\mu_{1\pm}$ and
$\mu_{2\pm}(\zeta)\equiv\mu_{2\pm}$ is as follows
\begin{eqnarray}\nonumber
&\pm&|\vec{\mu}_{\pm}|^{\lambda-7} +
\frac{\lambda-1}{4}|\vec{\mu}_{\pm}|^4 +
\frac{c}{2b}\mu_{1\pm}^2|\vec{\mu}_{\pm}|^2\\\label{e112} &+&
\frac{(\lambda-5)c}{4b}\mu_{1\pm}^2\mu_{2\pm}^2=0.
\end{eqnarray}
The observed dependence of Eqs.~(\ref{e111}), (\ref{e112}) on a
single variable $\zeta=h/\tau^{(\lambda-2)/(\lambda-3)}$ serves as
evidence of the validity of the scaling hypothesis. In a similar
way, the scaling for $h$, $\chi_{\parallel}$ and $\chi_{\perp}$ may
be confirmed.

\section{Conclusions}

Although second order phase transitions take place only under
certain conditions, e.g. $\tau=0$, $h=0$ for magnetic systems, the
singular part of the free energy and its thermodynamic derivatives
are described by functions characterized by scaling properties
nearby. Such properties are found experimentally by measurements at
various points in temperature and small values of the field.
Introducing appropriate scaling fields, data collapse to universal
scaling functions is used to determine the exponents. On the other
hand, logarithmic corrections present in the temperature and field
dependence of quantities described otherwise by power laws may
disturb this collapse. Therefore we have on the one hand calculated
several scaling functions for ranges the network parameter $\lambda$
where power laws are valid and found the characteristic dependence
of these functions on $\lambda$ even when the exponents are
independent of $\lambda$. On the other hand we have determined the
possible logarithmic corrections at certain borderline values of
$\lambda$.

It is remarkable that already within mean field theory such
logarithmic corrections arise. They are attributed  to correlations
due to the network properties rather than fluctuations (which are
absent in mean field) of the spin properties themselves. However we
note that an essential assumption in the mean field treatment was
the additive contribution of the individual nodes of the same degree
to the network free energy.

In this paper, we gave a comprehensive description of the
temperature and field behavior of a system with two coupled scalar
order parameters on a scale-free network in the vicinity of the
second order phase transition point. Special attention has been paid
to the appearance of the logarithmic corrections to scaling. For
magnetic systems on $d$-dimensional lattices, such behavior arises
due to the order parameter fluctuations that tend to be strongly
correlated in the vicinity of the critical point. It is the space
dimension $d$ that is definitive for the relevance of such
fluctuations. The scale-free networks we consider here are not
characterized by the Euclidean metrics and the space dimension.
Instead, it is the node degree distribution function exponent
$\lambda$ (\ref{rrr001}) that brings about correlations present in a
network due to its internal structure. As $\lambda$ decreases and
the node degree distribution becomes more and more fat-tailed, the
relative number of the high-degree nodes (hubs) increases and leads
to non-trivial critical behavior. This can be related to the fact
that below $\lambda=5$ the fourth moment of the degree distribution
diverges and below $\lambda=3$ the second moment ceases to exist.
First, for $\lambda_c=5$, the non-trivial dependencies appear. With
further decrease of $\lambda$, for $\lambda<3$, the systems appears
to be ordered at any finite temperature.

One observes a certain formal similarity between the behavior of
spin systems on $d$-dimensional lattices and on scale-free networks
with exponent $\lambda$. Both at $d=d_c$ and at  $\lambda=\lambda_c$
the logarithmic corrections to scaling are precursors of the change
in the critical behavior. This formal similarity is further
pronounced in a more subtle way: the correction exponents found by
us for the scale-free networks, although numerically different from
those found on the lattices
\cite{ON_log,ONlr_log,phi3_log,Potts_log,Shalaev94}, obey the same
scaling relations. In the present work we derived two new scaling
relations for logarithmic correction exponents (\ref{newSR1}) and
(\ref{newSR2}). Together with previously found \cite{Kenna06}
relations (\ref{rrr018}), (\ref{rrr019}) they form a complete set of
scaling relations for  logarithmic corrections.

\section*{Acknowledgement}

We acknowledge useful discussions with Bertrand Berche (Nancy), Ihor
Mryglod (Lviv), and J\.{o}zef Sznajd (Wroc{\l}aw). This work was
supported by the Austrian Fonds zur F\"orderung der
wissenschaftlichen Forschung under Project No. P19583-N20 (R.F. and
Yu.H.) and by the Austrian-Ukrainian Bureau of Cooperation in
Science, Education, and Culture (V.P.). It is our special pleasure
to thank the {\em Groupe de Physique Statistique} (Nancy University)
for their wonderful hospitality during MECO 35 (Pont-\`{a}-Mousson)
where this paper has been finalized.

\begin{appendix}
\renewcommand{\thesection}{\Alph{section}}%
\renewcommand{\theequation}{\Alph{section}.\arabic{equation}}%
\section{}
In this Appendix we give the deviation of the exponents governing
the field dependencies of thermodynamic quantities at $\tau=0$ for
different values of $\lambda$.

\subsection{Case $\lambda>5$}

In this case the free energy follows (\ref{feg5}) and the system of
equations of state reads:
\begin{equation}\label{eq49}
a(T-T_c)x_1 + bx_1|\vec{x}|^2 + \frac{1}{2}cx_1x_2^2 = h,
\end{equation}
\begin{equation}\label{eq50}
a(T-T_c)x_2 + bx_2|\vec{x}|^2 + \frac{1}{2}cx_1^2x_2 = 0.
\end{equation}
At $\tau=0$ one finds two different solutions, which we now detail.

{\bf Solution $x_1\neq0$, $x_2=0$.} In the case of vanishing $x_2$
this solution is
\begin{equation}
x_1 = \frac{1}{b^{1/3}}h^{1/3}, \hspace{2em} x_2=0.
\end{equation}
This solution exists and satisfies the stability conditions if
\begin{equation}\label{s1}
b>0, \hspace{2em} c>-2b.
\end{equation}
The longitudinal and transverse susceptibilities follow
\begin{equation}
\chi_\parallel = \frac{1}{3b^{1/3}}h^{-2/3},
\end{equation}
\begin{equation}
\chi_\perp = \frac{2b^{2/3}}{2b+c}h^{-2/3}.
\end{equation}
The heat capacity at the critical point reads
\begin{equation}
C_h = \frac{a^2}{3b}T_c.
\end{equation}

{\bf Solution $x_1\neq0$, $x_2\neq0$.} For $x_2$ non-vanishing

\begin{equation}
x_1 = \Big[-\frac{4b}{c(4b+c)}\Big]^{1/3} h^{1/3}, \hspace{1em}
x_2=\pm\sqrt{-\frac{2b+c}{2b}}x_1.
\end{equation}
Note, that the ratio between $x_1$ and $x_2$ does not depend on the
strength of the field and depends only on the ratio $c/b$. This
solution exists and satisfies stability conditions at
\begin{equation}\label{s2}
b>0, \hspace{2em} -4b<c<-2b.
\end{equation}
The susceptibilities read
\begin{equation}
\chi_\parallel = -\frac{8b}{c(8b+c)
-\sqrt{\xi}}\Big[\frac{c(4b+c)}{4b}\Big]^{2/3}h^{-2/3},
\end{equation}
\begin{equation}
\chi_\perp = -\frac{8b}{c(8b+c)
+\sqrt{\xi}}\Big[\frac{c(4b+c)}{4b}\Big]^{2/3}h^{-2/3},
\end{equation}
where
\begin{equation}\label{a6}
\xi = c^2(8b+c)^2 - 48bc(2b+c)(4b+c).
\end{equation}
The heat capacity follows:
\begin{equation}
C_h = \frac{2}{3}\frac{a^2}{(4b+c)}T_c.
\end{equation}

From the above solutions we conclusion that the exponents defined in
formulas (\ref{rrr005})--(\ref{rrr007}) are
\begin{equation}\label{a7}
\delta=3,\hspace{2em}\gamma_c=\frac{2}{3}, \hspace{2em} \alpha_c=0.
\end{equation}

\subsection{Case $\lambda=5$}

Given the free energy (\ref{fe5}), the system of equations of state
reads
\begin{eqnarray}\label{a10}
a(T-T_c)x_1 + bx_1|\vec{x}|^2\ln\frac{1}{|\vec{x}|} -
\frac{b}{4}x_1|\vec{x}|^2\\\nonumber +
\frac{c}{2}x_1x_2^2\ln\frac{1}{|\vec{x}|} -
\frac{c}{4}\frac{x_1^3x_2^2}{|\vec{x}|^2} = h,
\end{eqnarray}
\begin{eqnarray}\label{a11}
a(T-T_c)x_2 + bx_2|\vec{x}|^2\ln\frac{1}{|\vec{x}|} -
\frac{b}{4}x_2|\vec{x}|^2\\\nonumber +
\frac{c}{2}x_1^2x_2\ln\frac{1}{|\vec{x}|} -
\frac{c}{4}\frac{x_1^2x_2^3}{|\vec{x}|^2} = 0.
\end{eqnarray}
Note that this is a system of transcendent equations and one may
estimate the solution at weak external field $h\to0$. At $\tau=0$
there exist two solutions of the system (\ref{a10}), (\ref{a11}).

{\bf Solution $x_1\neq0$, $x_2=0$.} The solution for vanishing $x_2$
is
\begin{equation}
x_1\approx \Big(\frac{3}{b}\Big)^{1/3}\frac{h^{1/3}}{(-\ln
h)^{1/3}},\hspace{2em} x_2=0.
\end{equation}
This solution exists and satisfies the stability conditions if
\begin{equation}
b>0, \hspace{2em} c>-2b.
\end{equation}
The susceptibilities follow
\begin{equation}
\chi_\parallel = (9b)^{-1/3}h^{-2/3}(-\ln h)^{-1/3},
\end{equation}
\begin{equation}
\chi_\perp =
\frac{2b}{2b+c}\Big(\frac{b}{3}\Big)^{-1/3}h^{-2/3}(-\ln h)^{-1/3}.
\end{equation}
The heat capacity reads
\begin{equation}
C_H = \frac{a^2}{b}T_c(-\ln h)^{-1}.
\end{equation}

{\bf Solution $x_1\neq0$, $x_2\neq0$.} For nonzero $x_2$ , the
solution is of the form
\begin{eqnarray}
x_1 &\approx&
\Big(-\frac{c}{2b}\Big)^{2/3}\Big(\frac{6}{4b+c}\Big)^{1/3}h^{1/3}(-\ln
h)^{-1/3},\\\nonumber x_2 &=& \pm\sqrt{-\frac{2b+c}{2b}}x_1.
\end{eqnarray}
This solution exists and satisfies stability conditions if
\begin{equation}
b>0, \hspace{2em} -4b<c<-2b.
\end{equation}
The susceptibilities follow
\begin{equation}
\chi_\parallel = \chi^{II}_\parallel h^{-2/3}(-\ln h)^{-1/3},
\end{equation}
\begin{equation}
\chi_\perp = \chi^{II}_\perp h^{-2/3}(-\ln h)^{-1/3},
\end{equation}
where
\begin{equation}
\chi^{II}_\parallel = -\frac{6^{1/3}4b}{c(8b+c) -\sqrt{\xi}}
\Big(-\frac{c}{2b}\Big)^{5/3}(4b+c)^{2/3},
\end{equation}
\begin{equation}
\chi^{II}_\perp = -\frac{6^{1/3}4b}{c(8b+c) +\sqrt{\xi}}
\Big(-\frac{c}{2b}\Big)^{5/3}(4b+c)^{2/3},
\end{equation}
and $\xi$ is defined by Eq.(\ref{a6}). The heat capacity is
\begin{equation}
C_h = \frac{2a^2}{4b+c}T_c(-\ln h)^{-1}.
\end{equation}

Comparing the obtained solutions with the definition of the
logarithmic corrections to scaling exponents
(\ref{rrr012})--(\ref{rrr017}), we arrive at
\begin{equation}\label{corr}
\hat{\delta}=-\frac{1}{3},\hspace{2em}\hat{\gamma_c}=-\frac{1}{3},
\hspace{2em} \hat{\alpha_c}=-1.
\end{equation}

\subsection{Case $3<\lambda<5$}

The system with the free energy (\ref{fe35}) is described by the
following equations of state
\begin{eqnarray}\nonumber
a(T-T_c)x_1 &+& \frac{\lambda-1}{4}bx_1|\vec{x}|^{\lambda-3} +
\frac{1}{2}cx_1x_2^2|\vec{x}|^{\lambda-5}\\\label{a20}
&+&\frac{\lambda-5}{4}cx_1^3x_2^2|\vec{x}|^{\lambda-7} = h,
\end{eqnarray}
\begin{eqnarray}\nonumber
a(T-T_c)x_2 &+& \frac{\lambda-1}{4}bx_2|\vec{x}|^{\lambda-3} +
\frac{1}{2}cx_1^2x_2|\vec{x}|^{\lambda-5}\\\label{a21}
&+&\frac{\lambda-5}{4}cx_1^2x_2^3|\vec{x}|^{\lambda-7} = 0.
\end{eqnarray}

The system of equations (\ref{a20}), (\ref{a21}) has two solutions.

{\bf Solution $x_1\neq0$, $x_2=0$.} For the vanishing $x_2$ this
solution is
\begin{equation}
x_1 =
\Big(\frac{4}{(\lambda-1)b}\Big)^{1/(\lambda-2)}h^{1/(\lambda-2)},
\hspace{2em} x_2=0
\end{equation}
This solution exists and satisfies the stability conditions if
\begin{equation}
b>0, \hspace{2em} c>-\frac{\lambda-1}{2}b.
\end{equation}
The longitudinal and transverse susceptibilities read
\begin{equation}
\chi_\parallel =
\frac{1}{\lambda-2}\Big(\frac{4}{(\lambda-1)b}\Big)^{1/(\lambda-2)}h^{-(\lambda-3)/(\lambda-2)},
\end{equation}
\begin{equation}
\chi_\perp =
\frac{4}{(\lambda-1)b+2c}\Big(\frac{(\lambda-1)b}{4}\Big)^{(\lambda-3)/(\lambda-2)}h^{-(\lambda-3)/(\lambda-2)}.
\end{equation}
The heat capacity
\begin{equation}
C_h =
\frac{a^2}{\lambda-2}\Big(\frac{4}{(\lambda-1)b}\Big)^{3/(\lambda-2)}T_c
h^{(5-\lambda)/(\lambda-2)}.
\end{equation}

{\bf Solution $x_1\neq0$, $x_2\neq0$.} For the non-vanishing $x_2$
the solution is
\begin{widetext}

\begin{equation}
x_1 = \Big[\frac{4(1+\mu^2)^{(7-\lambda)/2}}{(\lambda-1)b(1+\mu^2)^2
+ 2c\mu^2(1+\mu^2) - (5-\lambda)c\mu^2} \Big]^{1/(\lambda-2)}
h^{1/(\lambda-2)},\hspace{2em} x_2 = \pm\mu x_1,
\end{equation}
where
\begin{equation}
\mu^2 = \frac{-2(\lambda-1)b - (\lambda-3)c + \sqrt{(\lambda-3)^2c^2
- 4(\lambda-1)(5-\lambda)bc}}{2(\lambda-1)b}.
\end{equation}

The solution exists and is stable if
\begin{equation}
b>0, \hspace{2em} -4b<c<-\frac{\lambda-1}{2}b.
\end{equation}
The susceptibilities follow
\begin{equation}
\chi_{\parallel,\perp} = \frac{\Big[(\lambda-1)b(1+\mu^2)^2 +
2c\mu^2(1+\mu^2) +
(\lambda-5)c\mu^2\Big]^{(\lambda-3)/(\lambda-2)}}{\Big\{(\lambda-1)^2b
+ 2c\Big\}(1+\mu^2)^3 - (\lambda+3)(5-\lambda)c\mu^2(1+\mu^2) \pm
\sqrt{D} }
4^{1/(\lambda-2)}2(1+\mu^2)^{\frac{\lambda+3}{2(\lambda-2)}}h^{-(\lambda-3)/(\lambda-2)},
\end{equation}
where
\begin{eqnarray}
D = (1-\mu^2)^2\Big\{[(\lambda-1)(\lambda-3)b - 2c](1-\mu^2)^2 +
(5-\lambda)(7-\lambda)c\mu^2\Big\}^2\\
+ 4\mu^2\Big\{[(\lambda-1)b + 2c](\lambda-3)(1+\mu^2)^2 +
(5-\lambda)(7-\lambda)c\mu^2\Big\}^2.
\end{eqnarray}
Finally, the heat capacity is
\begin{equation}
C_h = \Big[\frac{4(1+\mu^2)^{(7-\lambda)/2}}{(\lambda-1)b(1+\mu^2)^2
+ 2c\mu^2(1+\mu^2) +
(\lambda-5)c\mu^2}\Big]^{3/(\lambda-2)}\frac{1+\mu^2}{\lambda-2}a^2T_c
h^{(5-\lambda)/(\lambda-2)}.
\end{equation}
\end{widetext}
The above results give the following values of the critical
exponents:
\begin{equation}\label{exps}
\delta=\lambda-2,\hspace{2em}\gamma_c=\frac{\lambda-3}{\lambda-2},
\hspace{2em} \alpha_c=\frac{\lambda-5}{\lambda-2}.
\end{equation}
Note that in this case all leading exponents that govern the field
dependencies at $\tau=0$ are $\lambda$-dependent.

Summarizing results of Ref. \cite{Palchykov09}, Eqs.
(\ref{exp1})--(\ref{exp3}), and those obtained in this section, Eqs.
(\ref{a7}), (\ref{corr}), and (\ref{exps}) we give the values of the
leading and correction to scaling exponents in Tables \ref{tab1},
\ref{tab2} completing them by the gap exponents $\Delta$,
$\hat{\Delta}$ calculated via Eqs. (\ref{gap}).

\end{appendix}


\begin{thebibliography}{50}
\bibitem{note1} If the Helmholtz potential $F(\tau,m)$
is a generalized homogeneous function, one can show \cite{Hankey72}
that the other thermodynamic potentials are also generalized
homogeneous functions.
\end{thebibliography}
\end{document}